\documentclass[11pt]{article}
\usepackage{amsmath,amssymb,amsthm,bm}
\usepackage[authoryear]{natbib}
\usepackage{bbm}
\usepackage{tikz}
\usepackage{multirow}
\usepackage{graphicx}
\usepackage[T1]{fontenc}
\usepackage{calligra}
\usepackage{subcaption}
\usepackage[margin=.9in,hmargin=1in]{geometry}
\usepackage{setspace}
\newtheorem{thm}{Theorem}

\newtheorem{lemma}{Lemma}
\newtheorem{prop}{Proposition}

\newcommand{\mean}{\mathbb E}

\newcommand{\bs}{\boldsymbol}
\newcommand{\mb}{\mathbf}

\providecommand{\keywords}[1]
{
  \small	
  \textbf{\textit{Keywords---}} #1
}
\title{Joint Mean-Vector and Var-Matrix estimation\\
for Locally Stationary VAR(1) processes}
\author{Giovanni Motta\footnote{Correpondence to: Giovanni Motta, 3143 TAMU, Department of Statistics, College Station, TX 77843, USA.
E-mail: g.motta@stat.tamu.edu}\\
Department of Statistics, Texas A\&M University}
\date{}

\usepackage{fancyhdr}
\makeatletter
\let\runauthor\@author
\let\runtitle\@title
\makeatother
\begin{document}
\maketitle
\setcounter{page}{0}
\thispagestyle{empty}
\begin{abstract}
During the last two decades, locally stationary processes have been widely studied in the time series literature. In this paper we consider the locally-stationary vector-auto-regression model of order one, or LS-VAR(1), and estimate its parameters by weighted least squares. The LS-VAR(1) we consider allows for a smoothly time-varying non-diagonal VAR matrix, as well as for a smoothly time-varying non-zero mean. The weighting scheme is based on kernel smoothers. The time-varying mean and the time-varying VAR matrix are estimated {\it jointly}, and the definition of the local-linear weighting matrix is provided in closed-from. The quality of the estimated curves is illustrated through simulation results.

\end{abstract}
\qquad\keywords{Local Stationarity, Local Polynomials, Weighted Least Squares}

\vfill
\noindent Data Availability Statement:
Data sharing not applicable to this article as no datasets were generated or analyzed during the current study.

\newpage

\section{Introduction}

In this paper we consider $r$-dimensional multivariate data $\bs X_T(1),\dots \bs X_T(t)$ generated by a locally stationary process, and our goal is to fit to the data a parametric model with time-varying coefficients. The notation $\bs X_T(t)$ emphasizes that the data is a triangular array where at each $t$, the structure of the process depends on the sample size $T$.

To introduce the problem, consider the following uni-variate zero-mean autoregressive model of order $p$,
\begin{equation}\label{model0}
\sum_{j=0}^p a_j(\tfrac t T) X_T(t-j)=\sigma(\tfrac t T)\varepsilon(t),
\end{equation}
or AR($p$), where the coefficients $a_j(u)$ are differentiable for $u\in(0,1)$ with bounded derivatives. 

In terms of modeling, local stationarity means that if the parameters $a_j(u)$ are smooth in rescaled time $u$ and $T$ is large, then $a_j(\tfrac s T)\approx a_j(\tfrac t T)$ for all $s$ in a neightbour of $t$. For estimation, rescaling in time allows to apply non-parametric methods to recover the unknown curves. In the frequency domain, the importance of rescaling time $t$ by the sample size $T$ and developing the analysis in rescaled time $u\in(0,1)$ relies upon the uniqueness of the transfer function. \cite{D96} introduced 
the spectral representation of a locally stationary process
\[
X_T(u) = \int_{-\pi}^{-\pi}\exp(i\lambda t) A_T^0(t,\lambda)d\xi(\lambda),
\]
where $\xi(\lambda)$ is a stochastic process with orthogonal increments, and where the sequence $A_T^0(t,\lambda)$ converges (uniformly in $t$ and $\lambda$, as $T$ diverges) to another function $A(u,\lambda)$:
\[
\sup_{t,\lambda} |A_T^0(t,\lambda) - A(\tfrac t T,\lambda)|=\mathcal O(\tfrac 1 T).
\]
If $A$ is smooth in $u$, then the time-varying spectral density $f(u,\lambda)=| A(\tfrac t T,\lambda)|^2$ is uniquely determined from the triangular array. 

The dichotomy between $A_T^0$ and $A$ is particularly relevant in the case of AR processes. To see this, consider the simple case where $p=1$ and and $\sigma(u)\equiv 1$. In the stationary case where the coefficient $a$ is time-invariant, the process in \eqref{model0} can be written as
\[
X(t)=\sum_{k=0}^\infty \psi_{\,\!_k} \varepsilon_{t-k}
\]
with $\psi_{\,\!_k}=a^k$. By contrast, the locally stationary process $X_T(t)=a(\tfrac t T) X_T(t-1)+\varepsilon(t)$ does not have a solution of the form 
\[
X_T(t)=\sum_{k=0}^\infty \psi_{\,\!_k}(u) \varepsilon_{t-k},
\]
but only of the form $
X_T(t)=\displaystyle\sum_{k=0}^\infty \psi_{\!_{T, k}}(t) \varepsilon_{t-k}$,\,\, with\,\, $
\sup_{t}\displaystyle\sum_{k\in\mathbb Z} |\psi_{\!_{T, k}}(t)- \psi_{\,\!_k}(\tfrac t T)|=\mathcal O(\tfrac 1 T)$.

The seminal papers on local stationarity \citep{D96,D97} provide details on the mathematics in the frequency domain. For an overview on multivariate locally stationary processes, see \citet[][Section 7.2]{D12}.

Without loss of generality, assume that $\sigma(u)$ is known and time-invariant, that is, $\sigma(u)\equiv\sigma$. Suppose that the vector of interest 
$\bs a(u)=[a_1(u),\dots,a_p(u)]^\top$ depends on a finite dimensional parameter. For example, if the coefficients are polynomials in time 
\begin{eqnarray}\label{gp}
\begin{split}
a_j(u)=&\sum_{k=1}^K \theta_{jk}f_k(u),\quad1\leq k\leq K,&\quad 1\leq j\leq p,\\
\mbox{with}\quad f_k(u)=&\,\,\,u^{k-1},\qquad\qquad 1\leq k\leq K,&
\end{split}
\end{eqnarray}
estimating the time-varying vector $\bs a(u)$ at $u\in(0,1)$ translates into estimating the time-invariant vector $
\bs\theta=[\bs\theta_1^\top,\dots,\bs \theta_p^\top]^\top$, where $\bs\theta_j=[\theta_{j1},\dots,\theta_{jK}]^\top$, with $1\leq j\leq p$. 

The specification in \eqref{gp} approximates the coefficients vector $\bs a(u)$ by global polynomials. \citet[][Section 4]{D97} obtained an explicit formula for the vector ${\bs\theta}$ as the solution of a linear system similar to the Yule-Walker equations. In the univariate setting, \cite{D96} estimates the time-varying parameters $\bs a(u)$ by kernel smoothers, that is, using a local-constant approximation. In the multivariate settings, \citet[][p. 1776]{D00} mentions the possibility of estimating the unknown parameters $\bs\theta (u)$ by minimazing of a local-polynomial approximation of the local likelihood.


\cite{ZW10} consider univariate linear models of the form $\bs Y(t)=\bs X^\top \bs\beta(t)+\bs\varepsilon(t)$, where both $\bs X$ and $\bs\varepsilon$ are asumed to be locally stationary, and estimate their time-varying vector of coefficients $\bs\beta(t)$ by means of local polynomials. 

The contribution of this paper is threefold. In terms of modeling, we consider a multivariate version of model \eqref{model0} and estimate the time varying parameters in time domain. Our main contribution is the closed-form definition of the local-linear estimator of the parameters. Finally, we emphasize that the estimation of time-varying mean and time-varying AR matrix is performed jointly.

In Section~\ref{sec:LSVAR} we derive the localized Yule-Walker equations for a locally-stationary zero-mean VAR process. In Section~\ref{sec:WLS} we  consider a locally-stationary VAR with time-varying mean. First, we derive the local-constant weighted-least-squares  estimator, see Proposition~\ref{prop:localco}. Then in Theorem~\ref{prop:localin} we establish our main result, the closed form definition of the local-linear weighted-least-squares estimator. In Section~\ref{sec:simres} we illustrate and compare the performance of the two weighted-least-squares estimators (local-constant and local-linear). Section~\ref{sec:concl} concludes and highlights the extension of our approach to the high-dimensional ($r>T$) setting.

Through the paper we use bold uppercase letters to denote matrices, and bold slanted to denote vectors. We denote by $\mathbf I_m$ the identity matrix of size $m$, by $\bs 1_m$ the $m\times 1$ vector of ones, by ${\rm tr}\{ \mathbf A\}$ the trace of $\mathbf A$, 
by $\mathbf A^\top$ the transpose of $\mathbf A$, by $\|\mathbf A\|$ the Frobenius norm $\|\mathbf A\|=[{\rm tr}\{ \mathbf A^\top \mathbf A\}]^{\scalebox{0.6}{$1/2$}}$, and by $\mathbf A^{-1}$ the inverse of $\mathbf A$, that is, the square matrix such that $\mathbf A^{-1}\mathbf A=\mathbf A\mathbf A^{-1}=\mathbf I$. Finally, we use the acronyms VAR and WLS for vector auto-regression and weighted least squares, respectively.

\section{Locally Stationary Vector Auto Regression}\label{sec:LSVAR}
Consider the following $r$-dimensional Locally Stationary Vector Auto Regression of order 1
\begin{equation}
\label{LSVAR1}
\underset{r\times 1}{\bs X_t} = \underset{r\times r}{\mathbf A(\tfrac t T)}\, \underset{r\times 1}{\bs X_{t-1}}+\underset{r\times 1}{ \bs\varepsilon _t},\qquad t=1,\dots, T,
\end{equation}
with $\bs X_0=\bs 0$ and $\mean[\bs\varepsilon_t \bs X_s^\top]=\mathbbm 1_{\{s=t\}}\mb\Gamma_\varepsilon$. 
If the largest eigenvalue of $\mb A$ lies inside the unit circle
\begin{equation*}
\sup_{u\in(0,1)}\left|{\rm v}_1[\mb A(u)]\,\right|< 1,
\end{equation*}
$\bs X_t$ is locally stationary and causal. Our goal is to estimate $\mathbf A(u)$ at a fixed $u\in(0,1)$ using a localized version of the Yule-Walker equations. If we assume that the matrix-valued function $\mb A(x)$ is smooth in $x$, we can write the Taylor expansion of $\mathbf A(x)$ around $u$:
\[
 \mathbf A(x)= \sum_{j=0}^\infty \tfrac{\big(x - u\big)^j}{j!} \mb A^{(j)}(u)\\
 				= \mb A(u)+(x -u)\mb A^{(1)}(u) +\mathcal O([x -u]^2)
\]
where $\mb A^{(j)}(u):=\tfrac{d^j \mb A(x)}{d x^j}|_{x=u}$. We are interested in evaluating the function $\mb A(\tfrac t T)$ at those value of $t$ in a neighborhood of $u$. For example, for a fixed $u_0\in(0,1)$ let  $t_0=\lfloor u_0 T\rfloor$, where $\lfloor x \rfloor$ is the largest integer not exceeding $x$.
 Then we have the following uniform bound:
\[
\sup_{u_0\in(0,1)} |\tfrac{t_0}T -u_0| <\tfrac 1 T.
\]
As a consequence, assuming that 
\[
\sup_{u\in(0,1)}\|\mb A^{(1)}(u)\| <\infty
\]
we obtain following bound
\[
\|\mathbf A(\tfrac t T)-\mb A(u)\| \leq |\tfrac t T- u|\,  \|\mb A^{(1)}(u)\| = \mathcal O(\tfrac 1 T)\times \mathcal O(1)=
 \mathcal O(\tfrac 1 T)
\]
uniformly in $u$. 
Let $K_h(x)=\tfrac 1 h K(\tfrac x h)$, where $K(\cdot)$ is a Kernel function such that $K(x/h)=0$ is $|x|>h$ and $h=h_{\!_T}$ is the smoothing parameter that tends to zero as $T$ diverges, but slower than $1/T$:
\[
h=h_{\!_T}\to 0\quad\mbox{and}\quad Th_{\!_T}\to\infty \quad\mbox{as}\, T\to\infty.
\]
If we right-multiply \eqref{LSVAR1} by $\bs X_{t-1}^\top K_h(\tfrac t T - u)$ and sum over $t$ we obtain

\begin{eqnarray*}
\sum_{t=1}^T \bs X_t\bs X_{t-1}^\top K_h(\tfrac t T - u)&=&\sum_{t=1}^T \mb A(\tfrac t T )\bs X_{t-1} \bs X_{t-1}^\top K_h(\tfrac t T - u)+\sum_{t=1}^T \bs\varepsilon _t\bs X_{t-1}^\top K_h(\tfrac t T - u).
\end{eqnarray*}
Since $\mean[\bs\varepsilon_t \bs X_{t-1}^\top]=0$, 
{\small
\[
\sum_{t=1}^T \mb A(\tfrac t T )\bs X_{t-1} \bs X_{t-1}^\top K_h(\tfrac t T - u)
-
\mb A(u)\sum_{t=1}^T \bs X_{t-1} \bs X_{t-1}^\top K_h(\tfrac t T - u)=
\sum_{t=1}^T [\mb A(\tfrac t T )-\mb A(u)]
\bs X_{t-1} \bs X_{t-1}^\top K_h(\tfrac t T - u),
\]}
\vspace{-.5cm}
\begin{eqnarray*}
\|\tfrac 1 T\sum_{t=1}^T \bs X_{t} \bs X_{t}^\top K_h(\tfrac t T - u) - 
\mb\Gamma(0,u)\|=\mathcal O_p(\tfrac 1{\sqrt{T\, h}}),\\
\|\tfrac 1 T\sum_{t=1}^T \bs X_{t} \bs X_{t-1}^\top K_h(\tfrac t T - u) - 
\mb\Gamma(1,u)\|=\mathcal O_p(\tfrac 1{\sqrt{T\, h}}),
\end{eqnarray*}
and 
{\small
\begin{eqnarray*}
\|
\tfrac 1 T\sum_{t=1}^T  [\mathbf A(\tfrac t T)-\mb A(u)]
 \bs X_{t-1}\bs X_{t-1}^\top K_h(\tfrac t T - u)\|&\leq&
\sup_{t,T}\|\mathbf A(\tfrac t T)\|  \times
\|\tfrac 1 T
\sum_{t=1}^T  \bs X_{t-1}\bs X_{t-1}^\top K_h(\tfrac t T - u)
\|\\
& \leq& |\tfrac t T- u|\, \times \|\mb A^{(1)}(u)\| \times\|\tfrac 1 T\sum_{t=1}^T \bs X_{t-1}\bs X_{t-1}^\top K_h(\tfrac t T - u)\| \\
& \leq&\mathcal O(\tfrac 1 T)\times \mathcal O(1)\times\|\mb\Gamma(0,u)+\mathcal O_p(\tfrac 1{\sqrt{T\, h}})\|=\mathcal O_p(\tfrac 1 T),
\end{eqnarray*}}
it makes sense estimating $\mb A(u)$ as
\begin{equation}\label{Aest}
\widehat{\mb A}(u) =\Big[\sum_{t=1}^T \bs X_{t} \bs X_{t-1}^\top K_h(\tfrac t T - u)\Big]
\Big[\sum_{t=1}^T \bs X_{t-1} \bs X_{t-1}^\top K_h(\tfrac t T - u)\Big]^{-1}.
\end{equation}
If we define
\begin{eqnarray}
\underset{T\times r}{\mathbf X_0}&=&\{\underset{r\times 1}{\bs X_0},\underset{r\times 1}{\bs X_1}, \dots,\underset{r\times 1}{\bs X_{T-1}}\}^\top,\label{X0}\\
\underset{T\times r}{\mathbf X_1}&=&\{\underset{r\times 1}{\bs X_1},\underset{r\times 1}{\bs X_2}, \dots,\underset{r\times 1}{\bs X_T}\}^\top,\,\,\mbox{and}\label{X1}\\
\underset{T\times T}{\mathbf K_{\!_T}(u)}&=&{\rm{diag}}\{K_h(\tfrac 1 T - u),K_h( \tfrac 2 T - u),\dots,K_h(1 -u)\},\label{WTu}
\end{eqnarray}
the estimator in \eqref{Aest} can be written as
\begin{equation}\label{AestW}
\widehat{\mb A}(u) = [\mathbf X_1^\top \mathbf K_{\!_T}(u) \mathbf X_0]\,  [\mathbf X_0^\top \mathbf K_{\!_T}(u) \mathbf X_0]^{-1}.
\end{equation}

\section{Joint estimation of time-varying mean-vector and VAR-matrix by Weighted Least Squares}\label{sec:WLS}
The estimator in \eqref{AestW} can be obtained as the minimizer of a WLS problem. Consider the LS-VAR(1) in \eqref{LSVAR1}, and let us now allow for a time-varying (non-zero) mean
\begin{equation}\label{lsVARm}
\bs X_t -\bs\mu(\tfrac t T)=\mb A(\tfrac t T)[\bs X_{t-1} -\bs\mu(\tfrac{t-1} T)] + \bs\varepsilon_t,\quad t=1,\dots,T,
\end{equation}
with $\bs X_0=\bs\mu(0)=\bs 0$ and $\mean[\bs\varepsilon_t \bs X_s^\top]=\mathbbm 1_{\{s=t\}}\mb\Gamma_\varepsilon$. 
If the largest eigenvalue of $\mb A$ lies inside the unit circle
\begin{equation}\label{supvAm}
\sup_{u\in(0,1)}\left|{\rm v}_1[\mb A(u)]\,\right|< 1,
\end{equation}
$\bs X_t$ is locally stationary and causal. Our goal is to estimate $\mathbf A(u)$ at a fixed $u\in(0,1)$ by WLS. We can rewrite \eqref{lsVARm} as
\begin{equation}\label{lsVARm2}
\bs X_t= \bs m(\tfrac t T)+ \mb A(\tfrac t T)\bs X_{t-1} + \bs\varepsilon_t,
\end{equation}
where $\bs m(\tfrac t T)=\bs\mu(\tfrac t T) -\mb A(\tfrac t T)\bs\mu(\tfrac{t-1} T)$ for all $t=1,\dots,T$. 
If we define 
\begin{eqnarray*}
\underset{r\times(r+1)}{\mb B(u)}&=&[\bs m(u),\mb A(u)]\quad\mbox{and}\\
\underset{(r+1)\times 1}{\bs Z_t}&=&
\begin{pmatrix}
1\\
\bs X_{t-1}
\end{pmatrix}
\end{eqnarray*}
for all $t=1,\dots,T$, model \eqref{lsVARm2} can be written as
\begin{equation}\label{model}
\bs X_t= \mb B(\tfrac t T)\bs Z_t + \bs\varepsilon_t,
\end{equation}
and it makes sense to define $\widehat{\mb B}(u)$ as the minimizer of the weighted loss function
\begin{equation}\label{wls}
\sum_{t=1}^T \|\bs X_t-\mb B(\tfrac t T)\bs Z_t\|^2\,K_h(\tfrac t T -u),
\end{equation}
where the bandwidth sequence $h\equiv h_T$ tends to zero slower than $T^{-1}$: $h_T\to 0$ and $T\,h_T\to\infty$ as $T\to\infty$. The following proposition provides a closed-form of the local-constant minimizer of \eqref{wls}.
We consider model \eqref{model} and use the local-constant approximation of $\mathbf B(\tfrac t T)$ in a neighborhood 
of $u$
\[
\mathbf B(\tfrac t T)\approx\mathbf B(u)
\]
to estimate $\mathbf B(u)$, our parameter of interest, in the approximate model
\[
\bs X_t \approx \mathbf B(u) \bs Z_t +  \bs\varepsilon_t.
\]
Our first result generalizes the Yule-Walker solutions in \eqref{AestW} to allow for the time-varying mean vector $\bs\mu$.

\begin{prop}\label{prop:localco}

Let $\bs X_t$ follow the locally stationary model in \eqref{lsVARm}, with $\mb A(u)$ satisfying \eqref{supvAm}. Assume that the mean function $\bs\mu(u)$ and the VAR matrix $\mb A(u)$ in \eqref{lsVARm} are both differentiable uniformly in $u$, that is,
\begin{equation}
\sup_{u\in(0,1)} \|\bs\mu^{(1)}(u)\|<\infty, \qquad \mbox{and}\quad
\sup_{u\in(0,1)} \|\mb A^{(1)}(u)\|<\infty,\label{deriv1}
\end{equation}
where $\bs\mu^{(1)}(u):=\tfrac{d \bs\mu(x)}{d x}|_{x=u}$ and $\mb A^{(1)}(u):=\tfrac{d \mb A(x)}{d x}|_{x=u}$. Then, the local constant minimizer of \eqref{wls} is 
\begin{eqnarray}
\widehat{\mb B}(u)&=&\mb X_1^\top\mathbf K_{\!_T}(u)\mb Z_0\,\big(\mb Z_0^\top \mathbf K_{\!_T}(u)\mb Z_0\big)^{-1}\label{prop1A}\\ &=&[\widehat{\bs m}(u),\widehat{\mb A}(u)],\label{prop1B}
\end{eqnarray}
with $\widehat{\mb A}(u) = \widehat{\mb G}(u,1)\, [\widehat{\mb G}(u,0)^{-1}]$\,\, and\,\, 
$\widehat{\bs m}(u)=\widehat{\bs\mu}_1(u) - \widehat{\mb A}(u)\widehat{\bs\mu}_0(u)$,
where 
\begin{align}
\widehat{\bs\mu}_0(u)&=& \mb X_0^\top \mathbf K_{\!_T}(u)\bs 1_T/\bs 1_T^\top \mathbf K_{\!_T}(u) \bs 1_T\quad=&\sum_{t=1}^T\tfrac {K_h(\frac t T -u)}{\sum_{s=1}^T K_h(\frac s T -u)} \bs X_{t-1},\label{mu0}\\
\widehat{\bs\mu}_1(u)&=& \mb X_1^\top \mathbf K_{\!_T}(u)\bs 1_T/\bs 1_T^\top \mathbf K_{\!_T}(u) \bs 1_T\quad=&\sum_{t=1}^{T}\tfrac {K_h(\frac t T -u)}{\sum_{s=1}^T K_h(\frac s T -u)} \bs X_t,\label{mu1}\\
\widehat{\mb G}(u,0)&=&[\mb X_0^\top\mathbf K_{\!_T}(u)\mb X_0- \tfrac{\mb X_0^\top\mathbf K_{\!_T}(u)\bs 1_T\bs 1_T^\top\mathbf K_{\!_T}(u)\mb X_0}{\bs 1_T^\top \mathbf K_{\!_T}(u) \bs 1_T}]= &[\mb X_0 - \bs 1_T\widehat{\bs\mu}_0^\top(u)]^\top\mathbf K_{\!_T}(u)[\mb X_0 - \bs 1_T\widehat{\bs\mu}_0^\top(u)]\label{G0}\\
&=&[\mb X_0 - \tfrac{\bs 1_T\bs 1_T^\top\mathbf K_{\!_T}(u)\mb X_0}{\bs 1_T^\top \mathbf K_{\!_T}(u) \bs 1_T}]^\top
\mathbf K_{\!_T}(u)[\mb X_0 - \tfrac{\bs 1_T\bs 1_T^\top\mathbf K_{\!_T}(u)\mb X_0}{\bs 1_T^\top \mathbf K_{\!_T}(u) \bs 1_T}]
=&\sum_{t=1}^T[\bs X_{t-1} - \widehat{\bs\mu}_0(u)][\bs X_{t-1} - \widehat{\bs\mu}_0(u)]^\top K_h(\tfrac t T -u),\nonumber\\
\widehat{\mb G}(u,1)&=&[\mb X_1^\top\mathbf K_{\!_T}(u)\mb X_0- \tfrac{\mb X_1^\top\mathbf K_{\!_T}(u)\bs 1_T\bs 1_T^\top\mathbf K_{\!_T}(u)\mb X_0}{\bs 1_T^\top \mathbf K_{\!_T}(u) \bs 1_T}]=&[\mb X_1 - \bs 1_T\widehat{\bs\mu}_1^\top(u)]^\top\mathbf K_{\!_T}(u)[\mb X_0 - \bs 1_T\widehat{\bs\mu}_0^\top(u)]\label{G1}\\
&=&
[\mb X_1 - \tfrac{\bs 1_T\bs 1_T^\top\mathbf K_{\!_T}(u)\mb X_1}{\bs 1_T^\top \mathbf K_{\!_T}(u) \bs 1_T}]^\top
\mathbf K_{\!_T}(u)
[\mb X_0 - \tfrac{\bs 1_T\bs 1_T^\top\mathbf K_{\!_T}(u)\mb X_0}{\bs 1_T^\top \mathbf K_{\!_T}(u) \bs 1_T}]
=&\sum_{t=1}^T[\bs X_t - \widehat{\bs\mu}_1(u)][\bs X_{t-1} - \widehat{\bs\mu}_0(u)]^\top K_h(\tfrac t T -u).\nonumber
\end{align}

\begin{proof}
See Appendix~\ref{app:localco}.
\end{proof}
\end{prop}

The following proposition provides a closed-form of the local-linear minimizer of \eqref{wls}. We consider model \eqref{model} and use the local-linear approximation of $\mathbf B(\tfrac t T)$ in a neighborhood 
of $u$
\begin{equation}\label{Bapprox}
\mathbf B(\tfrac t T)\approx\mathbf B(u)+(\tfrac t T -u)\mathbf B^{(1)}(u)
\end{equation}
to estimate $\mathbf B(u)$, our parameter of interest, in the approximate model
\begin{equation}\label{Xapprox}
\bs X_t \approx \mathbf B(u) \bs Z_t +(\tfrac t T - u) \mathbf B^{(1)}(u) \bs Z_t+  \bs\varepsilon_t.
\end{equation}
\begin{thm}\label{prop:localin}

Let $\bs X_t$ follow the locally stationary model in \eqref{lsVARm}, with $\mb A(u)$ satisfying \eqref{supvAm}. Assume that the mean function $\bs\mu(u)$ and the VAR matrix $\mb A(u)$ in \eqref{lsVARm} are both continuously differentiable, uniformly in $u$, that is,
\begin{equation}
\sup_{u\in(0,1)} \|\bs\mu^{(2)}(u)\|<\infty, \qquad \mbox{and}\quad
\sup_{u\in(0,1)} \|\mb A^{(2)}(u)\|<\infty,\label{deriv2}
\end{equation}
where $\mb A^{(2)}(u):=\tfrac{d^2 \mb A(x)}{d x^2}|_{x=u}$.  Let
\[
\underset{T\times (r+1)}{\mb Z_0}=[\mb 1_T|\mb X_0],
\]
where $\mb X_0$ has been defined in \eqref{X0}, and define the diagonal matrix
\begin{equation}\label{D1u}
\underset{T\times T}{\mb\Delta_1(u)}={\rm{diag}}\{\tfrac t T -u,\,1\leq t\leq T\},
\end{equation}
and the weighting matrix 
\begin{equation}\label{WTuX}
\mb W_{\!_T}(u; \mb X)=  \mb K_T(u)- \mb K_T(u)\mb\Delta_1(u)\mb Z_0
[\mb Z_0^\top \mb \Delta_1(u)\mb K_T(u)\mb\Delta_1(u)\mb Z_0]^{-1}
\mb Z_0^\top \mb\Delta_1(u) \mb K_T(u),
\end{equation}
where $ \mb K_T(u)$ has been defined \eqref{WTu}. Then, the local linear minimizer of \eqref{wls} is 
\begin{eqnarray}
\widetilde{\mb B}(u)&=& 
\mb X_1^\top \mb W_{\!_T}(u; \mb X) \mb Z_0\,[{\mb Z}_0^\top \mb W_{\!_T}(u; \mb X){\mb Z}_0]^{-1}\label{prop2A}\\
&=&[\widetilde{\bs m}(u),\widetilde{\mb A}(u)],\label{prop2B}
\end{eqnarray}
with $\widetilde{\mb A}(u) = \widetilde{\mb G}(u,1)\, [\widetilde{\mb G}(u,0)^{-1}]$\,\,and\,\,
$\widetilde{\bs m}(u) = \widetilde{\bs\mu}_1(u) - \widetilde{\mb A}(u)\widetilde{\bs\mu}_0(u)$,
where 
\begin{eqnarray}
\widetilde{\bs\mu}_0(u)&=& \mb X_0^\top \mb W_{\!_T}(u; \mb X)\bs 1_T/\bs 1_T^\top \mb W_{\!_T}(u; \mb X) \bs 1_T\label{mu0tilde}\\
\widetilde{\bs\mu}_1(u)&=& \mb X_1^\top \mb W_{\!_T}(u; \mb X)\bs 1_T/\bs 1_T^\top \mb W_{\!_T}(u; \mb X) \bs 1_T\label{mu1tilde}\\
\widetilde{\mb G}(u,0)&=&[\mb X_0^\top\mb W_{\!_T}(u; \mb X)\mb X_0- \tfrac{\mb X_0^\top\mb W_{\!_T}(u; \mb X)\bs 1_T\bs 1_T^\top\mb W_{\!_T}(u; \mb X)\mb X_0}{\bs 1_T^\top \mb W_{\!_T}(u; \mb X) \bs 1_T}]\label{G0tilde}\\
\widetilde{\mb G}(u,1)
&=&[\mb X_1^\top\mb W_{\!_T}(u; \mb X)\mb X_0- \tfrac{\mb X_1^\top\mb W_{\!_T}(u; \mb X)\bs 1_T\bs 1_T^\top\mb W_{\!_T}(u; \mb X)\mb X_0}{\bs 1_T^\top \mb W_{\!_T}(u; \mb X) \bs 1_T}]\label{G1tilde},
\end{eqnarray}
$\mb X_1$ being defined in \eqref{X1}.

\begin{proof}
See Appendix~\ref{app:localin}.
\end{proof}
\end{thm}

\section{Simulation Results}\label{sec:simres}
We first consider the zero-mean locally stationary VAR(1) model in \eqref{LSVAR1}, and estimate the VAR matrix by means of the localized Yule-Walker equations. Then we consider the locally stationary VAR(1) model in \eqref{lsVARm} with time-varying mean, and compare the WLS estimates obtained with local-constant and local-linear weights, respectively.

\begin{figure}[h]
\centering
\begin{subfigure}[t]{0.45\linewidth}
\centering
\includegraphics[width=3.4in]{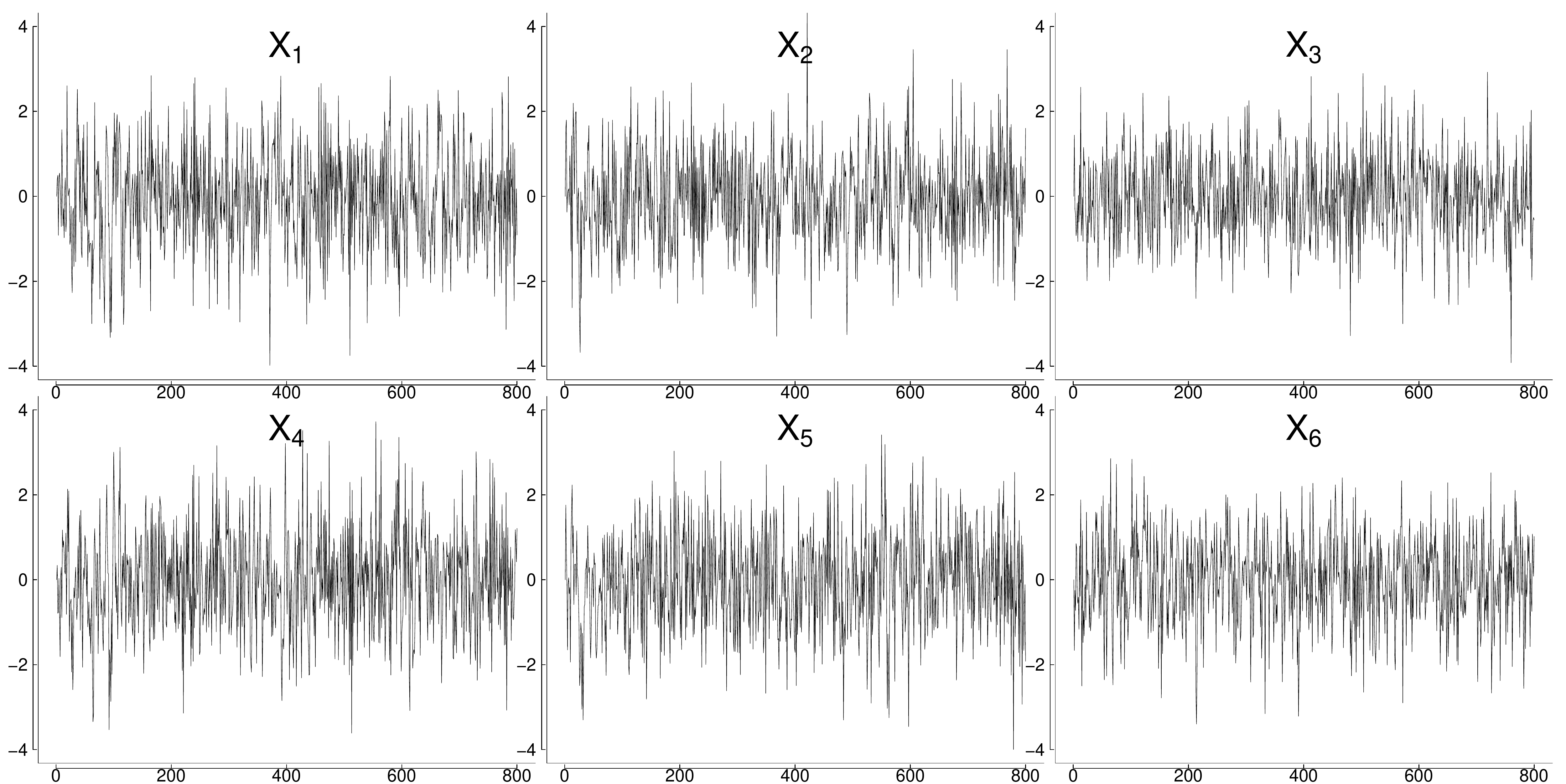}
\caption{\sf One realization of the $r=6$ times series simulated according to \eqref{LSVAR1} with $\mb A(u)$ as in \eqref{TVAR1sim} and $T=800$.}
\label{fig:est-factors}
\end{subfigure}\hspace{1.5cm}
\begin{subfigure}[t]{0.45\linewidth}
\centering
\includegraphics[width=3in]{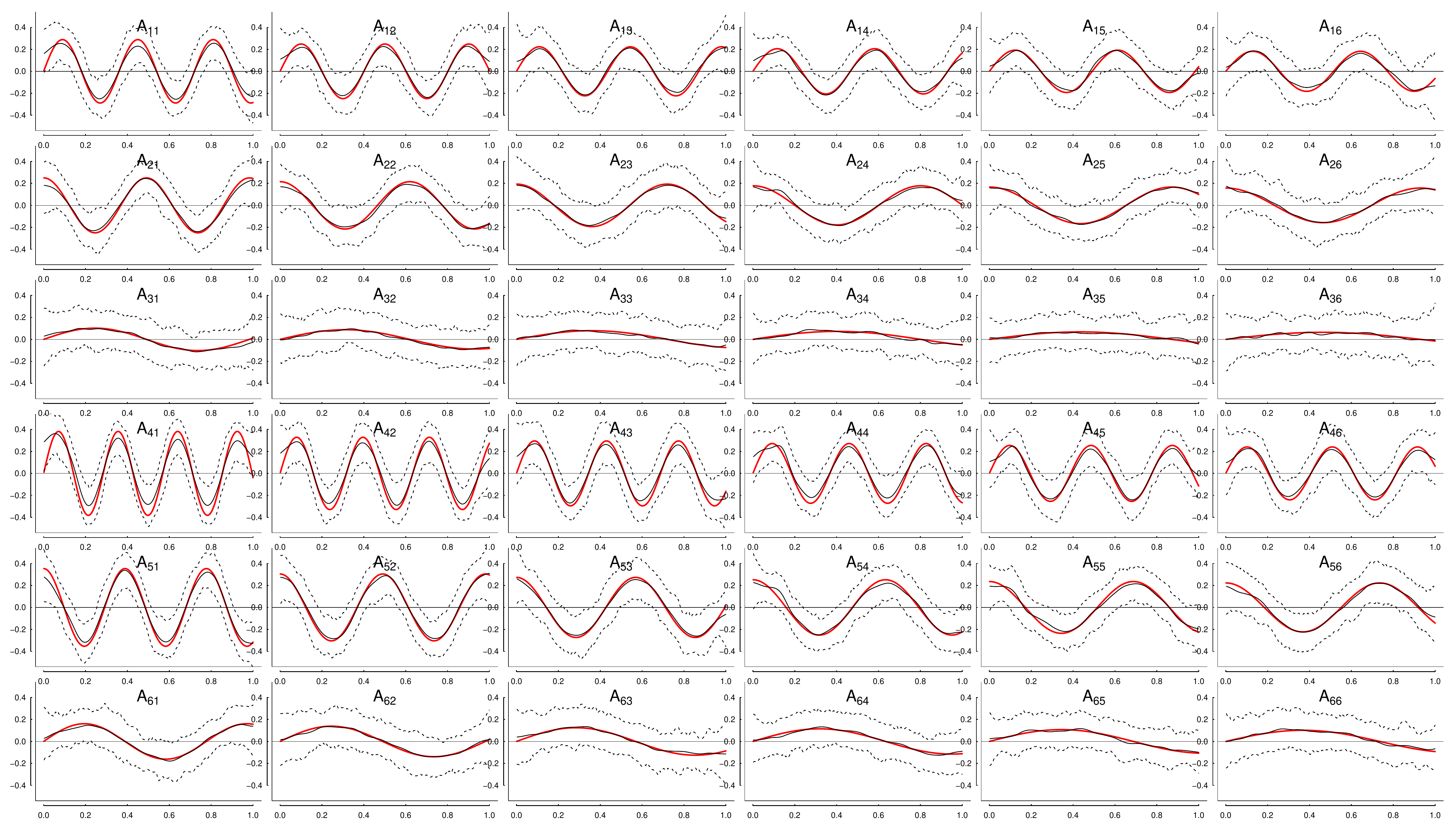}
\caption{\sf 
Red lines: simulated parameters according to \eqref{TVAR1sim}. Solid-black lines: average of the estimates over $M$ replications.
Dashed-black lines: 90\% confidence bands (empirical quantiles) of the $M$ estimates.}
\label{fig:est-A}
\end{subfigure}
\caption{\sf Left: simulated time series according to \eqref{LSVAR1}, with $r=6$ and $\mb A(u)$ as in \eqref{TVAR1sim}. Right: estimates $\widehat{\mb A}(u)$ obtained according to \eqref{AestW} over $M=100$ replications.}
\label{fig:simVAR}
\end{figure}

We simulate model \eqref{LSVAR1} with $r=6$. For $j=1,4$ and $k=1,\dots,r$, we generate
the time varying entries of the $r\times r$ matrix $\mb A(\tfrac tT)$ as
\begin{eqnarray}
    A_{j,k}(\tfrac tT)&=&a_1\tfrac{\sqrt{j+3}}{\log(k+3)}\sin\big(4\pi \tfrac t T \tfrac{\sqrt{j+4}}{\log(k+4)}\big),\nonumber\\
    A_{j+1,k}(\tfrac tT)&=&a_1\tfrac{\sqrt{j+2}}{\log(k+3)}\cos\big(2\pi \tfrac t T \tfrac{\sqrt{j+4}}{\log(k+2)}\big),\label{TVAR1sim}\\
    A_{j+2,k}(\tfrac tT)&=&a_2\tfrac{\sqrt{j+1}}{\log(k+3)}
		\sin\big(\pi \tfrac t T \tfrac{\sqrt{j+4}}{\log(k+2)}\big),\nonumber
\end{eqnarray}
with $T=800$, $a_1=0.2$ and $a_2=0.1$. This specification is such that, for all $u\in(0,1)$,
$$0.1<\left|{\rm v}_1[\mb A(u)]\,\right|< 0.9.$$
We estimate the parameters according to \eqref{AestW}, with $h=0.03$ and the  Gaussian Kernel $K(x)=\tfrac 1{\sqrt{2\pi}}\exp(-0.5\,x^2)$. 
The results are reported in Figure~\ref{fig:simVAR}.

\begin{figure}
\centering
\begin{subfigure}[t!]{0.45\linewidth}
\centering
\includegraphics[width=3.1in]{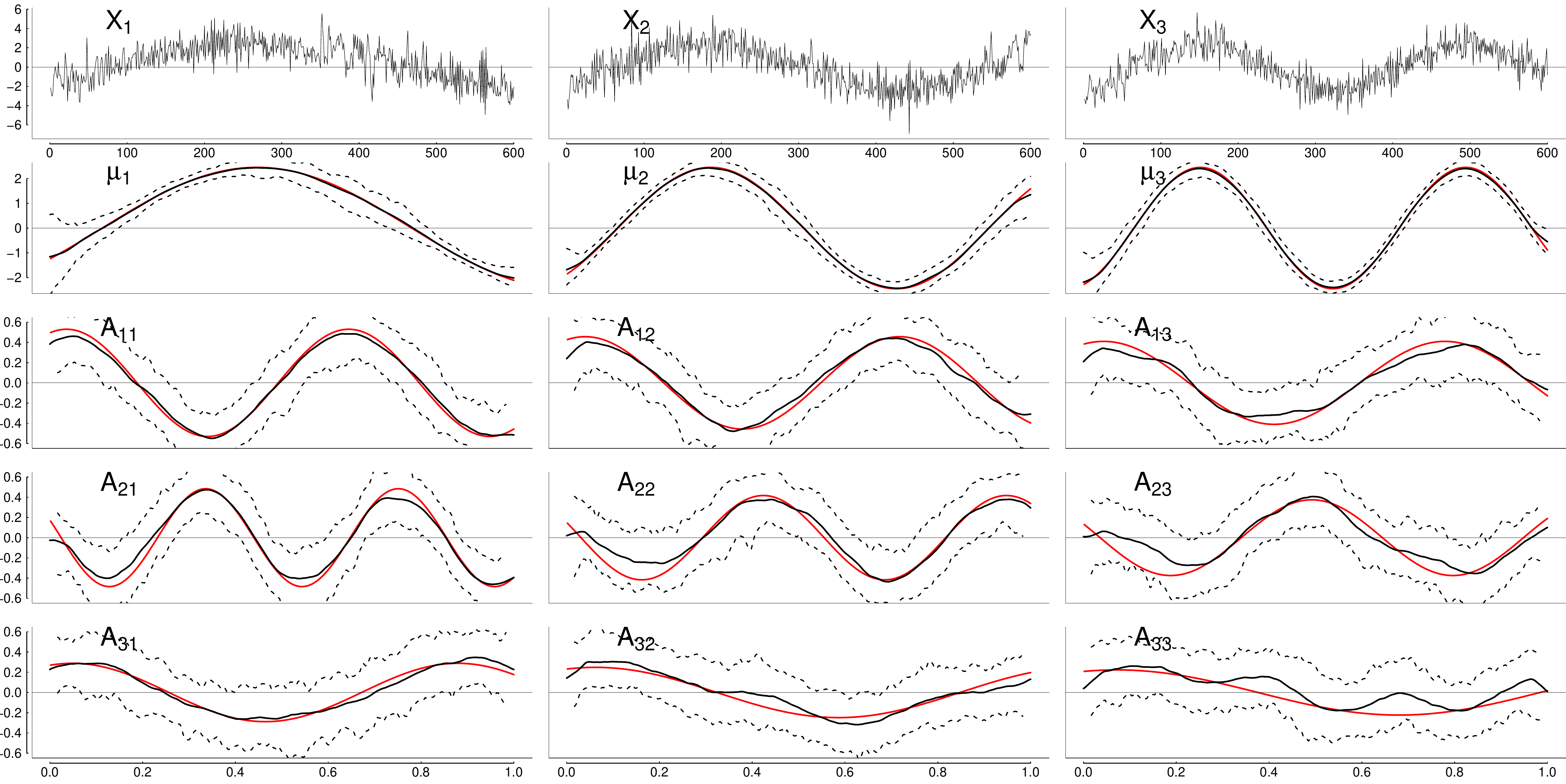}
\caption{\sf Local-constant WLS estimates obtained according to \eqref{prop1A}-\eqref{prop1B} over $M=100$ replications.
}
\label{fig:simVAR-LC}
\end{subfigure}\qquad
\begin{subfigure}{0.45\linewidth}
\includegraphics[width=3.1in]{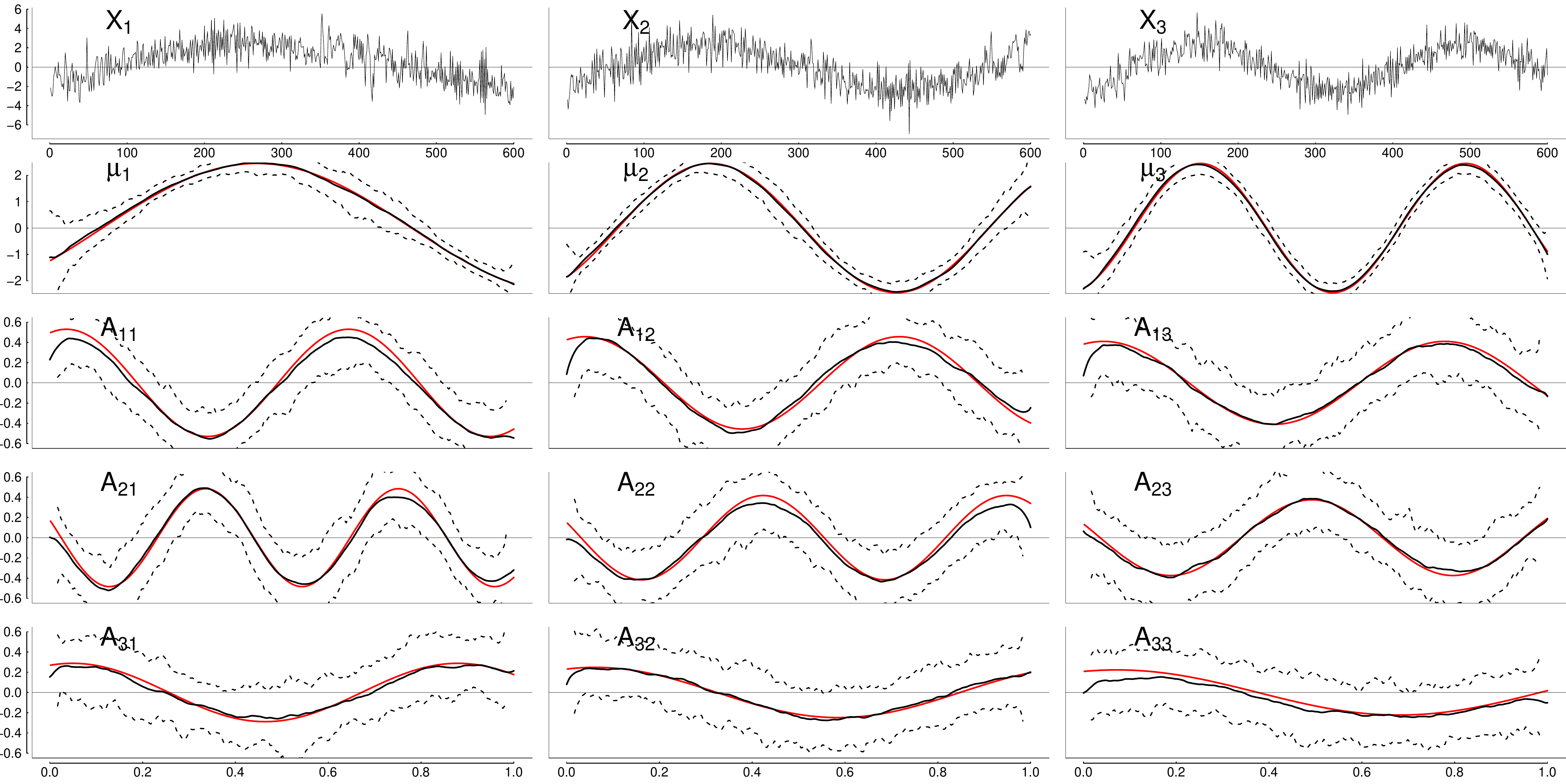}
\caption{\sf Local-linear WLS estimates obtained according to \eqref{WTuX}-\eqref{prop2B} over $M=100$ replications.}
\label{fig:simVAR-LL}
\end{subfigure}
\caption{
\sf First row: one realization of the $r=3$ time series simulated according to \eqref{lsVARm} with $\bs\mu(u)$ as in \eqref{mu1sim}, $\mb A(u)$ as in \eqref{VAR1sim}, and $T=600$. Second row: estimated time-varying means. Third, fourth, and last row: estimated time-varying VAR-coefficients. Red: simulated curves. Solid black: average of the estimates. Dashed black: 95\% confidence bands.} 
\label{fig:simVAR-LCLL}
\end{figure}

We simulate model \eqref{lsVARm} with $r=3$. For $k=1,\dots, r$, we generate
the time varying entries of the $r\times 1$ vector $\bs\mu(\tfrac tT)$ as
\begin{equation}\label{mu1sim}
\mu_k(\tfrac t T)=\sqrt{6}\,\sin(\pi\omega_k\,\tfrac t T-\phi_k)
\end{equation}
and the $r\times r$ matrix $\mb A(\tfrac tT)$ as
\begin{eqnarray}
A_{1,k}(\tfrac t T)&=&a_1\tfrac{\sqrt{6}}{\log(k+3)}\sin\big(1.2+2\pi \tfrac t T\tfrac{\sqrt{7}}{\log(k+4)}\big),\nonumber\\
A_{2,k}(\tfrac tT)&=&a_1\tfrac{\sqrt{5}}{\log(k+3)}\cos\big(1.2+ 2 \pi \tfrac t T \tfrac{\sqrt{7}}{\log(k+2)}\big),\label{VAR1sim}\\
A_{3,k}(\tfrac tT)&=&a_2\tfrac{\sqrt{4}}{\log(k+3)}\sin\big(1.2+\pi \tfrac t T \tfrac{\sqrt{7}}{\log(k+2)}\big),\nonumber
\end{eqnarray}
with $T=600$, $a_1=0.3$ and $a_2=0.2$, $\omega_k=0.5+ k$, and $\phi_k=0.2+ k/3$. This specification is such that, for all $u\in(0,1)$, $$0<\left|{\rm v}_1[\mb A(u)]\,\right|< 0.9.$$ Figure~\ref{fig:simVAR-LCLL} exhibits the parameters estimated by WLS with $h=0.04$ and the Epanechnikov Kernel $K(x)=\tfrac 3 4 (1-u^2)\mathbbm{1}_{\{|u|\leq 1\}}(x)$. The local constant estimates, obtained according to \eqref{prop1A}-\eqref{prop1B}, are presented reported in Figure~\ref{fig:simVAR-LC}.
The local linear estimates, obtained according to \eqref{WTuX}-\eqref{prop2B}, are reported in Figure~\ref{fig:simVAR-LL}.

Figure~\ref{fig:simVAR} shows that in the absence of the (time-varying) mean, that is when $\bs\mu(u)\equiv 0$, the local constant estimator performs very well. However, as Figure~\ref{fig:simVAR-LC} illustrates, this is not the case in the presence of a (time-varying) mean.

It is clear from Figure~\ref{fig:simVAR-LCLL} that 
although the local constant estimates look satisfactory, the local linear approach delivers superior results. As in the univariate case, the bias of the local linear estimator only depends on the second derivative (of the unknown regression function), and this does not come as a cost in the asymptotic variance. Moreover, the local linear estimator does not suffer from boundary bias problems. The quality of the estimates in Figure~\ref{fig:simVAR-LL} is remarkable.

\section{Conclusions and future research}\label{sec:concl}

In this paper we consider the problem of estimating the time-varying mean-vector and the time-varying AR-matrix of a locally stationary VAR(1). We provide the closed form definition of the local linear solution to the weighted least squares, in a way that the mean-vector and the AR-matrix are estimated jointly. 

The asymptotic properties of our estimator need to be studied. Also, it would be interesting to develop data driven methods to select the smoothing parameters. Moreover, we might consider the problem of estimating the parameters of a locally stationary VAR($p$) of order $p>1$. 

An  important contribution from future studies is the extension of the WLS in \eqref{wls5} to the high-dimensional setting: $r>>T$. To this end, the WLS approach can be can be generalized in more than one direction. In fact, the closed form in \eqref{WTuX}-\eqref{prop2B} of Theorem \ref{prop:localin} becomes particularly attractive in the case the length $r$ of the time series becomes large. Indeed, we can stick to the linear regression model in \eqref{Xapprox} 
with the same assumptions as in Section~\ref{sec:WLS}, and fit \eqref{Xapprox}  in a way to shrink the regression coefficients
towards zero. More precisely we can consider minimizing, with respect to $\mb B$, the following {\it{WLS-Ridge}} loss-function
\begin{equation}\label{WRR}
\|\mathbf X_1- \widetilde{\mathbf Z}_0 \mathbf B(u)^\top\|^2_{\mathbf K(u)}+\lambda
\|{\mathbf B}(u)\|^2,
\end{equation}
the effect of the penalty being to shrink the entries of $\mb B$ towards zero. The approach based on \eqref{WRR} can be generalized to the case of a non-spherical penalty. The loss function corresponding to this scenario is
\begin{equation}\label{GRR}
\|\mathbf X_1- \widetilde{\mathbf Z}_0{\mathbf B}(u)^\top\|^2_{\mathbf K(u)} +
\|{\mathbf B}(u)\|^2_{\mathbf\Lambda},
\end{equation}
which comprises a WLS criterion -- as in \eqref{wls5} -- and a generalized ridge penalty given by the matrix $\mathbf\Lambda(u)$. In both \eqref{WRR} and \eqref{GRR}, the  $(T\times T)$ matrix $\mathbf K(u)$ is diagonal, the $t$-th element $\{K_t(u)=\tfrac 1 h K(\tfrac{u- t/T}h),\, 1\leq t\leq T\}$ representing the weight of the $t$-th observation such that $\tfrac 1 T K_t(u)\in[0,1]$. The penalty in \eqref{GRR} is a quadratic form with penalty parameter $\mathbf\Lambda$, an $r$-dimensional positive-definite matrix. When  $\mathbf\Lambda=\lambda \mathbf I_r$, we obtain the spherical penalty of the WLS-Ridge  regression in \eqref{WRR}. Generalizing the (positive) scalar $\lambda$ to the class of (positive definite) matrices $\mb\Lambda$ allows for (i) different penalization per regression parameter, and (ii) joint shrinkage among the elements of $\mathbf B(u)$. 

\bibliographystyle{stat}
\bibliography{giovanni}

\newpage
\setcounter{page}{1}
\appendix
\section{Proof of Proposition~\ref{prop:localco}}
\label{app:localco}
If we assume that the matrix-valued function 
$$\mb B(x)=[\bs m(x),\mb A(x)]$$ is smooth in $x$, we can write the Taylor expansion of $\mathbf B(x)$ around $u$:
\begin{eqnarray*}
\mathbf B(x)&=&\sum_{j=0}^\infty\tfrac{(x-u)^j}{j!} \mb B^{(j)}(u)\\
 				&=& \mb B(u)+(x -u)\mb B^{(1)}(u) +\mathcal O([x -u]^2)
\end{eqnarray*}
where $\mb B^{(j)}(u):=\tfrac{d^j \mb B(x)}{d x^j}|_{x=u}$. Assuming \eqref{deriv1} implies that 
\[
\sup_{u\in(0,1)}\|\mb B^{(1)}(u)\| <\infty,
\]
and that
\[
\|\mathbf B(\tfrac t T)-\mb B(u)\| \leq |\tfrac t T- u|\,  \|\mb B^{(1)}(u)\| = \mathcal O(\tfrac 1 T)\times \mathcal O(1)=
 \mathcal O(\tfrac 1 T)
\]
uniformly in $u$, so that the loss in \eqref{wls} can be approximated by
\begin{equation}\label{wls2}
\sum_{t=1}^T \|\bs X_t-\mb B(u)\bs Z_t\|^2\,K_h(\tfrac t T -u).
\end{equation}
Letting
\[
\underset{T\times (r+1)}{\mb Z_0}=[\bs 1_T, \mb X_0]
\]
where $\bs 1_T$ is a $T\times 1$ vector of ones, and $\mb X_0$ has been defined in \eqref{X0}, the loss in \eqref{wls2} can be written in matrix form as
\begin{eqnarray}\label{wls3}
\begin{split}
\mathcal L_{\,\!_T}(u)&=\|\mb X_1 - \mb Z_0 \mb B(u)^\top\|^2_{\mb K_{\!_T}(u)}\\
&=
{\rm tr}\{[\mb X_1 - \mb Z_0 \mb B(u)^\top]^\top \,\mb K_{\!_T}(u)\, [\mb X_1 - \mb Z_0 \mb B(u)^\top]\},
\end{split}
\end{eqnarray}
with  $\mb X_1$ as in \eqref{X1} and $\mathbf K_{\!_T}(u)$ as in \eqref{WTu}. The loss in \eqref{wls3} is equal to
\begin{eqnarray*}
{\rm tr}\{[
\mb X_1^\top \mathbf K_{\!_T}(u) \mb X_1- 2\mb X_1^\top\mathbf K_{\!_T}(u)\mb Z_0 \mb B(u)^\top + 
\mb B(u)\mb Z_0^\top \mathbf K_{\!_T}(u)\mb Z_0\mb B(u)^\top ]\},
\end{eqnarray*}
and thus minimizing $\mathcal L_{\,\!_T}(u)$ with respect to $\mb B(u)$ is equivalent to minimizing 
\begin{eqnarray*}
{\rm tr}\{[\mb B(u)\mb Z_0^\top \mathbf K_{\!_T}(u)\mb Z_0\mb B(u)^\top 
- 2\mb X_1^\top\mathbf K_{\!_T}(u)\mb Z_0 \mb B(u)^\top]\}
\end{eqnarray*}
with respect to $\mb B(u)$. Differentiating w.r.t. $\mb B(u)^\top$ and equating to zero we obtain
\begin{eqnarray*}
{\rm tr}\{[2\mb B(u)\mb Z_0^\top \mathbf K_{\!_T}(u)\mb Z_0 - 2\mb X_1^\top\mathbf K_{\!_T}(u)\mb Z_0]\}=0,
\end{eqnarray*}
that is,
\begin{eqnarray*}
\widehat{\mb B}(u)=\mb X_1^\top\mathbf K_{\!_T}(u)\mb Z_0\,\big(\mb Z_0^\top \mathbf K_{\!_T}(u)\mb Z_0\big)^{-1},
\end{eqnarray*}
and thus \eqref{prop1A} is proved. Notice that 
\[
\underset{r\times(r+ 1)}{
\mb X_1^\top\mathbf K_{\!_T}(u)\mb Z_0}=
\big[
\underset{r\times 1}{\mb X_1^\top\mathbf K_{\!_T}(u)\bs 1_T} |
\underset{r\times r}{\mb X_1^\top \mathbf K_{\!_T}(u)\mb X_0}
\big],
\]
and that the matrix we need to invert can be partitioned as
\[
\underset{(r+1)\times(r+ 1)}{\mb Z_0^\top \mathbf K_{\!_T}(u)\mb Z_0}=
\left[
\begin{array}{c|c}
\underset{1 \times 1}{\bs 1_T^\top \mathbf K_{\!_T}(u) \bs 1_T}&
\underset{1\times r}{
\bs 1_T^\top \mathbf K_{\!_T}(u) \mb X_0}\\
\hline
\underset{r\times 1}{\mb X_0^\top \mathbf K_{\!_T}(u) \bs 1_T}&
\underset{r\times r}{\mb X_0^\top \mathbf K_{\!_T}(u) \mb X_0}
\end{array}
\right].
\]
Without proof we state the following Lemma, see 
\citet[][result (1)in Section 3.5.3, pages 29-30]{L}.
\begin{lemma}\label{partinv}
Let $\mb A$ be $m\times m$, $\mb B$ be $m\times n$, $\mb C$ be $n\times m$, and $\mb D$ be  $n\times n$, and consider the $(m+n)\times (m+n)$ partitioned matrix
\[
\left[
\begin{array}{c|c}
\mb A & \mb B\\
\hline
\mb C &\mb D
\end{array}
\right].
\]
If $\mb A$ and $[\mb D- \mb C \mb A^{-1}\mb B]$ are both nonsingular, then
\[
\left[
\begin{array}{c|c}
\mb A & \mb B\\
\hline
\mb C &\mb D
\end{array}
\right]^{-1}
=
\left[
\begin{array}{c|c}
\mb A^{-1} + \mb A^{-1} \mb B(\mb D- \mb C \mb A^{-1}\mb B)^{-1} \mb C \mb A^{-1}
& 
-\mb A^{-1}\mb B(\mb D- \mb C \mb A^{-1}\mb B)^{-1} \\
\hline
 -(\mb D- \mb C \mb A^{-1}\mb B)^{-1} \mb C \mb A^{-1}
&
(\mb D- \mb C \mb A^{-1}\mb B)^{-1}
\end{array}
\right].
\]
\end{lemma}
We can now prove \eqref{prop1B}, together with \eqref{mu0}, \eqref{mu1}, \eqref{G0} and \eqref{G1}.
By Lemma~\ref{partinv},
\[
\big(\mb Z_0^\top \mathbf K_{\!_T}(u)\mb Z_0\big)^{-1}=
\left[
\begin{array}{c|c}
(\bs 1_T^\top \mathbf K_{\!_T}(u) \bs 1_T)^{-1}+
\tfrac
{\bs 1^\top\mathbf K_{\!_T}(u)\mb X_0 
(\widehat{\mb G}(u,0)^{-1})
\mb X_0^\top\mathbf K_{\!_T}(u)\bs 1_T}{(\bs 1_T^\top \mathbf K_{\!_T}(u) \bs 1_T)^{2}}
&
-\tfrac{\bs 1^\top\mathbf K_{\!_T}(u)\mb X_0}{\bs 1^\top\mathbf K_{\!_T}(u)\bs 1_T}
(\widehat{\mb G}(u,0)^{-1})\\
\hline
-(\widehat{\mb G}(u,0)^{-1})\mb X_0^\top \mathbf K_{\!_T}(u) \bs 1_T
(\bs 1_T^\top \mathbf K_{\!_T}(u) \bs 1_T)^{-1}&
\widehat{\mb G}(u,0)^{-1}
\end{array}
\right],
\]
where $\widehat{\mb G}(u,0)$ has been defined in \eqref{G0}, 
and therefore
\[\widehat{\mb B}(u)=\mb X_1^\top\mathbf K_{\!_T}(u)\mb Z_0\,\big(\mb Z_0^\top \mathbf K_{\!_T}(u)\mb Z_0\big)^{-1}=[\widehat{\bs m}(u),\widehat{\mb A}(u)],\]
with 
\begin{eqnarray*}
\widehat{\mb A}(u)&=&-
\tfrac{\mb X_1^\top\mathbf K_{\!_T}(u)\bs 1_T\bs 1_T^\top\mathbf K_{\!_T}(u)\mb X_0}{\bs 1_T^\top\mathbf K_{\!_T}(u)\bs 1_T}
(\widehat{\mb G}(u,0)^{-1})
+
\mb X_1^\top\mathbf K_{\!_T}(u)\mb X_0 (\widehat{\mb G}(u,0)^{-1})\\
&=&
\widehat{\mb G}(u,1)\,\widehat{\mb G}(u,0)^{-1},
\end{eqnarray*}
where $\widehat{\mb G}(u,1)$ has been defined in \eqref{G1}, and
\begin{eqnarray*}
\widehat{\bs m}(u)&=&
\tfrac{\mb X_1^\top\mathbf K_{\!_T}(u)\bs 1_T}{\bs 1_T^\top\mathbf K_{\!_T}(u)\bs 1_T}+
\tfrac{\mb X_1^\top\mathbf K_{\!_T}(u)\bs 1_T\bs 1_T^\top \mathbf K_{\!_T}(u)\mb X_0 (\widehat{\mb G}(u,0)^{-1})\mb X_0^\top\mathbf K_{\!_T}(u)\bs 1_T
}{(\bs 1_T^\top\mathbf K_{\!_T}(u)\bs 1_T)^2}-
\tfrac{\mb X_1^\top\mathbf K_{\!_T}(u)\mb X_0 (\widehat{\mb G}(u,0)^{-1})\mb X_0^\top\mathbf K_{\!_T}(u)\bs 1_T}{\bs 1^\top\mathbf K_{\!_T}(u)\bs 1_T}\\
&=&
\tfrac{\mb X_1^\top\mathbf K_{\!_T}(u)\bs 1_T}{\bs 1_T^\top\mathbf K_{\!_T}(u)\bs 1_T}+
\big[\tfrac{\mb X_1^\top\mathbf K_{\!_T}(u)\bs 1_T\bs 1_T^\top \mathbf K_{\!_T}(u)\mb X_0}{\bs 1_T^\top\mathbf K_{\!_T}(u)\bs 1_T}-
\mb X_1^\top\mathbf K_{\!_T}(u)\mb X_0\big]
(\widehat{\mb G}(u,0)^{-1})\tfrac{\mb X_0^\top\mathbf K_{\!_T}(u)\bs 1_T}{\bs 1^\top\mathbf K_{\!_T}(u)\bs 1_T}\\
&=&
\widehat{\bs\mu}_1(u)-\widehat{\mb G}(u,1)\widehat{\mb G}(u,0)^{-1}\widehat{\bs\mu}_0(u)=
\widehat{\bs\mu}_1(u)-\widehat{\mb A}(u)\widehat{\bs\mu}_0(u),
\end{eqnarray*}
where $\widehat{\bs\mu}_0(u)$ and $\widehat{\bs\mu}_1(u)$ are given by \eqref{mu0} and \eqref{mu1}, respectively. 

\section{Proof of Theorem~\ref{prop:localin}}
\label{app:localin}
If we assume that the matrix-valued function 
$$\mb B(x)=[\bs m(x), \mb A(x)]$$
 is smooth in $x$, we can write the Taylor expansion of $\mathbf B(x)$ around $u$:
\begin{eqnarray*}
\mathbf B(x)&=& \sum_{j=0}^\infty \tfrac{(x- u)^j}{j!} \mb B^{(j)}(u)\\
 				&=& \mb B(u)+(x -u)\mb B^{(1)}(u) + \tfrac 1 2 (x -u)^2\mb B^{(2)}(u)
				+\mathcal O([x -u]^3)
\end{eqnarray*}
where $\mb B^{(j)}(u):=\tfrac{d^j \mb B(x)}{d x^j}|_{x=u}$. Assuming \eqref{deriv2} implies that 
\[
\sup_{u\in(0,1)}\|\mb B^{(2)}(u)\| <\infty,
\]
and that
\[
\left\|\mathbf B(\tfrac t T)-\big[\mb B(u)+(x -u)\mb B^{(1)}(u)\big]\right\| \leq \tfrac 1 2|\tfrac t T- u|^2\,  \|\mb B^{(2)}(u)\| = \mathcal O(\tfrac 1 {T^2})\times \mathcal O(1)=
 \mathcal O(\tfrac 1 {T^2})
\]
uniformly in $u$. Therefore, adopting \eqref{Bapprox}-\eqref{Xapprox} the loss in \eqref{wls} can be approximated by
\begin{equation}\label{wls4}
\sum_{t=1}^T \|\bs X_t-[\mb B(u)+(\tfrac t T- u)\mb B^{(1)}(u)]\bs Z_t\|^2\,K_h(\tfrac t T -u).
\end{equation}
Letting
\[
\underset{[2(r+1)]\times r}{\widetilde{\mb B}_1(u)^\top}=
\begin{bmatrix}
\underset{(r+1)\times r}{{\mb B}(u)^\top}\\
\underset{(r+1)\times r}{{\mb B}^{(1)}(u)^\top}
\end{bmatrix}
\]
and 
\begin{eqnarray*}
\underset{T\times [2(r+1)]}{\widetilde{\mb Z}_0}&=&[\mb Z_0|\mb\Delta_1(u)\,\mb Z_0],\quad\mbox{with}\\
\underset{T\times (r+1)}{\mb Z_0}&=&[\mb 1_T|\mb X_0],
\end{eqnarray*}
where $\bs 1_T$ is a $T\times 1$ vector of ones, and where $\mb X_0$ and $\mb\Delta_1(u)$ have been defined in \eqref{X0} and \eqref{D1u}, respectively, the loss in \eqref{wls4} can be written in matrix form as
\begin{eqnarray}\label{wls5}
\begin{split}
\widetilde{\mathcal L}_{\,\!_T}(u)&=\|\mb X_1 - \widetilde{\mb Z}_0 \widetilde{\mb B}_1(u)^\top\|^2_{\mb K_{\!_T}(u)}\\
&=
{\rm tr}\{[\mb X_1 - \widetilde{\mb Z}_0 \widetilde{\mb B}_1(u)^\top]^\top \,\mb K_{\!_T}(u)\, [\mb X_1 - \widetilde{\mb Z}_0 \widetilde{\mb B}_1(u)^\top]\},
\end{split}
\end{eqnarray}
with  $\mb X_1$ as in \eqref{X1} and $\mb K_{\!_T}(u)$ as in \eqref{WTu}. The loss in \eqref{wls5} is equal to
\begin{eqnarray*}
{\rm tr}\{[
\mb X_1^\top \mb K_{\!_T}(u) \mb X_1- 2\mb X_1^\top\mb K_{\!_T}(u)\widetilde{\mb Z}_0\widetilde{\mb B}_1(u)^\top + 
\widetilde{\mb B}_1(u)\widetilde{\mb Z}_0^\top \mb K_{\!_T}(u)\widetilde{\mb Z}_0\mb B_1(u)^\top ]\},
\end{eqnarray*}
and thus minimizing $\widetilde{\mathcal L}_{\,\!_T}(u)$ with respect to $\widetilde{\mb B}_1(u)$ is equivalent to minimizing 
\begin{eqnarray*}
{\rm tr}\{[\widetilde{\mb B}_1(u)\widetilde{\mb Z}_0^\top \mb K_{\!_T}(u)\widetilde{\mb Z}_0\widetilde{\mb B}_1(u)^\top 
- 2\mb X_1^\top\mb K_{\!_T}(u)\widetilde{\mb Z}_0 \widetilde{\mb B}_1(u)^\top]\}
\end{eqnarray*}
with respect to $\widetilde{\mb B}_1(u)$. Differentiating w.r.t. $\widetilde{\mb B}_1(u)^\top$ and equating to zero we obtain
\begin{eqnarray*}
{\rm tr}\{[2\,\widetilde{\mb B}_1(u)\widetilde{\mb Z}_0^\top \mb K_{\!_T}(u)\widetilde{\mb Z}_0 - 2\mb X_1^\top\mb K_{\!_T}(u)\widetilde{\mb Z}_0]\}=0,
\end{eqnarray*}
that is,
\begin{eqnarray*}
\widetilde{\mb B}_1(u)=\mb X_1^\top\mb K_{\!_T}(u)\widetilde{\mb Z}_0\,\big(\widetilde{\mb Z}_0^\top \mb K_{\!_T}(u)\widetilde{\mb Z}_0\big)^{-1}.
\end{eqnarray*}
Notice that 
\[
\underset{r\times[2(r+ 1)]}{
\mb X_1^\top\mb K_{\!_T}(u)\widetilde{\mb Z}_0}=
\big[
\underset{r\times (r+1)}{\mb X_1^\top\mb K_{\!_T}(u)\mb Z_0} |
\underset{r\times (r+1)}{\mb X_1^\top\mb K_{\!_T}(u)\mb\Delta_1(u)\mb Z_0}
\big],
\]
and that
\begin{eqnarray*}
\underset{[2(r+1)]\times [2(r+1)]}{\widetilde{\mb Z}_0^\top \mb K_T(u)\widetilde{\mb Z}_0}
&=&
\left[\begin{array}{c|c}
\mb A&\mb B\\
\hline
\mb C&\mb D
\end{array}\right],
\end{eqnarray*}
where
\begin{eqnarray*}
\underset{(r+1)\times (r+1)}{\mb A}&=&\mb Z_0^\top \mb K_T(u)\mb Z_0\\
\underset{(r+1)\times (r+1)}{\mb B}&=&\mb Z_0^\top \mb K_T(u)\mb\Delta_1(u) \mb Z_0\\
\underset{(r+1)\times (r+1)}{\mb C}&=&\mb Z_0^\top \mb\Delta_1(u) \mb K_T(u)\mb Z_0\\
\underset{(r+1)\times (r+1)}{\mb D}&=&\mb Z_0^\top \mb \Delta_1(u)\mb K_T(u)\mb\Delta_1(u)\mb Z_0.
\end{eqnarray*}

The local-linear estimator $\widetilde{\mb B}(u)$ of ${\mb B}(u)=[\boldsymbol m(u),\mb A(u)]$ is given by the first $r+1$ columns of the $r\times [2(r+1)]$ matrix
\[
\underset{r\times [2(r+1)]}{\mb X_1^\top \mb K_T(u) \widetilde{\mb Z}_0}\,\,
\underset{[2\times (r+1)][2\times (r+1)]}{[\widetilde{\mb Z}_0^\top \mb K_T(u) \widetilde{\mb Z}_0]^{-1}}.
\]
Hence, we need the first $r+1$ columns of the $[2(r+1)]\times[2(r+1)]$ matrix 
$[\widetilde{\mb Z}_0^\top \mb K_T(u)\widetilde{\mb Z}_0]^{-1}$. Without proof we state the following Lemma, see 
\citet[][result (2) in Section 3.5.3, page 30]{L}.
\begin{lemma}\label{partinv2}
Let $\mb A$ be $m\times m$, $\mb B$ be $m\times n$, $\mb C$ be $n\times m$, and $\mb D$ be  $n\times n$, and consider the $(m+n)\times (m+n)$ partitioned matrix
\[
\left[
\begin{array}{c|c}
\mb A & \mb B\\
\hline
\mb C &\mb D
\end{array}
\right].
\]
If $\mb D$ and $[\mb A- \mb B \mb D^{-1}\mb C]$ are both nonsingular, then
\[
\left[
\begin{array}{c|c}
\mb A & \mb B\\
\hline
\mb C &\mb D
\end{array}
\right]^{-1}
=
\left[
\begin{array}{c|c}
[\mb A- \mb B \mb D^{-1}\mb C]^{-1} & 
-[\mb A- \mb B \mb D^{-1}\mb C]^{-1} \mb B\mb D^{-1}\\
\hline
-\mb D^{-1} \mb C [\mb A- \mb B \mb D^{-1}\mb C]^{-1}&
\mb D^{-1}+\mb D^{-1} \mb C[\mb A- \mb B \mb D^{-1}\mb C]^{-1} \mb B \mb D^{-1}
\end{array}
\right].
\]
\end{lemma}

Using Lemma~\ref{partinv2} we can write 
{\small\begin{eqnarray*}
\widetilde{\mb B}(u)=&
\mb X_1^\top \mb K_T(u) \mb Z_0 &
[\mb Z_0^\top \mb K_T(u)\mb Z_0 - \mb Z_0^\top\mb K_T(u)\mb\Delta_1(u)\mb Z_0\mb D^{-1}
\mb Z_0^\top \mb\Delta_1(u) \mb K_T(u)\mb Z_0
]^{-1}
\\&- 
\mb X_1^\top \mb K_T(u)\mb\Delta_1(u) \mb Z_0\mb D^{-1}\mb Z_0^\top \mb\Delta_1(u) \mb K_T(u)\mb Z_0 &
[\mb Z_0^\top \mb K_T(u)\mb Z_0 - \mb Z_0^\top\mb K_T(u)\mb\Delta_1(u)\mb Z_0\mb D^{-1}
\mb Z_0^\top \mb\Delta_1(u) \mb K_T(u)\mb Z_0
]^{-1},
\end{eqnarray*}}

\noindent where $\mb D=\mb Z_0^\top \mb \Delta_1(u)\mb K_T(u)\mb\Delta_1(u)\mb Z_0$. If we define the matrix $\mb W_{\!_T}(u; \mb X)$ according to \eqref{WTuX}, the local-linear estimator $\widetilde{\mb B}(u)$ can be written as
\[
\widetilde{\mb B}(u)= \mb X_1^\top \mb W_{\!_T}(u; \mb X) \mb Z_0\,
[{\mb Z}_0^\top 
\mb W_{\!_T}(u; \mb X)
{\mb Z}_0]^{-1},
\]
and \eqref{prop2A} is proved. The estimator $\widetilde{\mb B}(u)$ has the same form as the estimator $\widehat{\mb B}(u)$ in \eqref{prop1A} of Proposition~\ref{prop:localco}, with $\mb W_{\!_T}(u; \mb X)$ instead of $\mb K_{\!_T}(u)$. Hence the result in \eqref{prop2B} with 
$\widetilde{\bs\mu}_0(u)$, $\widetilde{\bs\mu}_1(u)$, $\widetilde{\mb G}(u,0)$, and $\widetilde{\mb G}(u,1)$ as in \eqref{mu0tilde}, \eqref{mu1tilde}, \eqref{G0tilde}, and \eqref{G1tilde}, respectively, follows directly from Proposition~\ref{prop:localco}. 

\end{document}